# AResNet-ViT: A Hybrid CNN-Transformer Network for Benign and Malignant Breast Nodule Classification in Ultrasound Images

ZHAO XIN*

School of Information Engineering, Dalian University, Dalian, Liaoning

ZHU QIANQIAN

School of Information Engineering, Dalian University, Dalian, Liaoning

WU JIALING

Department of Ultrasound, First Affiliated Hospital of Dalian Medical University, Dalian, Liaoning

**Abstract**

**Background** Breast cancer is one of the most common types of cancer in women worldwide. Early diagnosis is crucial for improving survival rates. In recent years, there has been extensive research on computer-aided diagnosis (CAD) using artificial intelligence techniques, specifically deep neural networks, to accurately and automatically predict the benign or malignant nature of breast nodules in ultrasound images. However, accurate classification faces challenges due to the similarity between breast lesions and surrounding tissues, as well as the overlapping appearances of partially benign and malignant nodules. To address these challenges, this study proposes a deep learning network that integrates convolutional neural networks (CNN) and Transformer by leveraging the complementary strengths of CNN in local feature extraction and Transformers in global feature extraction, aiming to enhance the classification of benign and malignant breast lesions in ultrasound images.

**Method** The proposed network AResNet-ViT adopts a dual-branch architecture for comprehensive local-global features extraction. In the local feature extraction branch, a residual network with multiple attention-guided modules is utilized. This design effectively captures local details and texture features specific to breast nodules, enhancing sensitivity to subtle changes within the nodules and aiding in accurate classification of their benign or malignant nature. The global feature extraction branch leverages a multi-head self-attention ViT network to capture the overall shape, boundary, and relationships with surrounding tissues. This enhances the understanding and modeling of both nodule-specific and global image features. The experiments were conducted on the publicly available BUSI dataset and the performance of the experiments was evaluated using metrics of ACC, TPR, TNR, and AUC. The binary Cross Entropy (BCE) loss function was applied in training process.

**Results** We conducted ablation experiments on the attention-guided design of the residual CNN branch. We also performed architecture ablation experiments on the CNN branch and the transformer branch separately, as well as using both branches together. Additionally, we compared the performance of our proposed AResNet-ViT network with that of classical classification models and the results from three recent papers published in the past three years. The experimental results demonstrate that the AResNet-ViT network, with its hybrid CNN-transformer structure and multi-attention mechanism, achieved the highest values for evaluation metrics including ACC, TPR, TNR, and AUC in both the ablation and comparison experiments. These values were 0.889, 0.861, 0.896, and 0.925, respectively.

**Conclusion** This study indicates that the fusion of CNN and Transformer networks can effectively improve the performance of the classification model, providing a robust and efficient solution for benign-malignant classification of breast nodules in ultrasound images.

## 1 INTRODUCTION

Breast nodules, which may manifest as cystic or solid masses, are frequently encountered in breast tissue and represent a prevalent condition among women. These nodules are categorized as either benign or malignant. Benign breast nodules do not pose a substantial health risk, whereas malignant breast nodules indicate the presence of cancerous proliferation, thereby posing a significant threat to women's overall physical and emotional well-being.

Regular breast screening, including mammography X-ray, breast ultrasound, and breast magnetic resonance imaging (MRI), plays a crucial role in early detection of breast nodules and diagnosis of breast cancer. Mammography X-ray imaging, despite its significant radiation exposure and limited imaging angles, is primarily employed for further screening of malignant nodules. MRI imaging, on the other hand, is time-consuming and expensive, making it impractical for routine outpatient examinations. Ultrasound imaging, with its advantages of radiation-free, affordability, convenience, speed,and versatility in imaging from various angles, has emerged as the primary modality for evaluating breast nodules [1]. Nevertheless, the diagnostic accuracy of ultrasound for breast nodules heavily relies on the clinical expertise of ultrasound practitioners. Consequently, variations in physician experience or the impact of visual fatigue often lead to misdiagnosis or missed diagnoses.

With the continuous advancement of artificial intelligence technology, researchers have extensively explored computer-aided diagnosis of ultrasound breast nodules. Their efforts focus on developing intelligent algorithms capable of automatically identifying and characterizing nodule areas in ultrasound images as benign or malignant. These algorithms leverage techniques such as deep learning and machine learning to train models for nodule identification and classification. This AI-assisted diagnostic approach holds the potential to enhance the accuracy and efficiency of ultrasound-based assessment of breast nodules, equipping clinicians with reliable auxiliary tools to support clinical decision-making and treatment planning. Conversely, certain malignant nodules may display clear borders and an aspect ratio smaller than 1, aligning with characteristics typically associated with benign nodules, which cause difficulty in AI identification.

In the past decade, deep learning-based approaches have achieved remarkable success in natural image classification and garnered widespread attention in the field of medical image recognition. Specifically, in the domain of ultrasound breast image classification and recognition, several studies have employed various CNN-based deep learning models to learn and extract features specific to breast nodules in ultrasound images. in 2016, Huynh et al. [2] used the ImageNet data set to preprocess VGGNet, ResNet and DenseNet, subsequently comparing the classification performance of these networks on breast ultrasound images. In 2017, Han et al. [3] utilized the GoogLeNet algorithm to discern benign/malignant ultrasonic breast nodules.

Byra et al. [4] introduced a matching layer into the pre-trained VGG19 network in 2018, aiming to enhance pixel intensity and improve the classification performance of breast nodules. In 2019, Chen Siwen et al. [5] employed the adaptive contrast enhancement (ACE) method for preprocessing and deployed the AlexNet model for the classification of benign and malignant breast nodules. Qi et al. [6] employed a deep convolutional neural network with multi-scale and skip connections to differentiate between ultrasound breast malignant nodules and solid benign nodules. In 2020, Zhuang et al. [7] utilized image decomposition to obtain blur-enhanced and bilaterally filtered images, enriching the input information of breast lesions and facilitating the classification of breast ultrasound images. Cao et al. [8] proposed a noise filter network (NFNet) for classifying nodules. They introduced dual softmax layers to address the issue of inaccurate labeling caused by manual labeling errors or data quality issues. In 2021, Kalaf et al. [9] used the VGG16 model with an attention mechanism to classify benign and malignant breast nodules, and combined binary cross entropy and hyperbolic cosine loss to improve classification performance. Saxena et al. [10] employed an enhanced dataset of 12,000 images to compare the performance of different methods in breast nodule classification. In 2022, Lu et al. [11] utilized a pre-trained ResNet18 with spatial attention and combined three distinct RNNs to predict the presence of benign and malignant breast nodules. Kang et al. [12] proposed a multi-branch network comprising a feature extraction sub-module, a classification sub-module, and a pixel attention sub-module to enhance the classification of benign and malignant breast nodules through attention mechanisms.

Although Deep convolutional neural networks (CNNs) have made significant progress in performance and effectiveness compared to traditional classification methods, CNNs are primarily suited for extracting local features and may struggle with extracting global features. In 2020, Dosovitskiy et al. [13] proposed the Vision Transformer (ViT) network, which utilizes a self-attention mechanism to extract global features, resulting in remarkable performance in image classification tasks. in 2021, Behnaz et al. [14] employed the ViT model for the classification of ultrasound breast nodules, achieving superior results compared to convolutional neural networks. This study underscores the effectiveness of the ViT model in learning global features for breast ultrasound image classification. Subsequently, other studies have also introduced modifications to the original ViT network, specifically tailored for the classification of breast ultrasound images[15-18], such as in 2023, Shareef, B. [18] proposed a hybrid multitask deep neural network, called Hybrid-MT-ESTAN, which combines CNNs and Swin Transformer to perform ultrasound breast tumor classification and segmentation.

The local features of ultrasound breast images capture the details and characteristics of nodules, while the global information and dependencies reflect the relationships and distinctions between nodules and surrounding tissues. To fully exploit the advantages of CNN in extracting local features and the capabilities of the Vision Transformer in extracting global features, this study proposes the integration of CNN and Vision Transformer to construct a classification network model. The main contributions of this work are summarized as:

(1) The proposed dual-branch network architecture, named AResNet-ViT, seamlessly integrates CNN and Transformer to harness local and global feature information, thereby significantly enhancing the performance of the classification model.

(2) The attention-guided residual network (AResNet) is designed for local feature extraction, aiming to capture the shape, texture, edges, and higher-level semantic features of nodules..

(3) The Vision Transformer (ViT) is utilized to capture the global dependencies among pixels in ultrasound images, enabling the generation of a comprehensive global feature representation for nodule images.

## 2 METHOD

The dual-branch architecture of AResNet-ViT, consisting of two branches, is shown in Figure 1. The upper branch of the network utilizes a residual network guided by multiple attentions to effectively capture the local details and texture features of breast nodules. This capability enhances sensitivity to subtle changes within the nodules, contributing to the accurate determination of benign and malignant nodules. On the other hand, the lower branch of the network utilizes a multi-head self-attention-based Vision Transformer (ViT) to capture the overall shape, boundary, and relationship between the nodule and the surrounding tissue and enhances the understanding of both the nodule and the overall image characteristics. By combining and encoding the features extracted from the local feature extraction branch and the global dependency feature extraction branch, the network can effectively utilize both local and global information to improve the accuracy of breast nodule classification. Each branch of the network outputs a one-dimensional feature, which is subsequently concatenated and then encoded by a fully connected multi-layer perceptron (MLP). Finally, the classification result is obtained through Sigmoid activation.

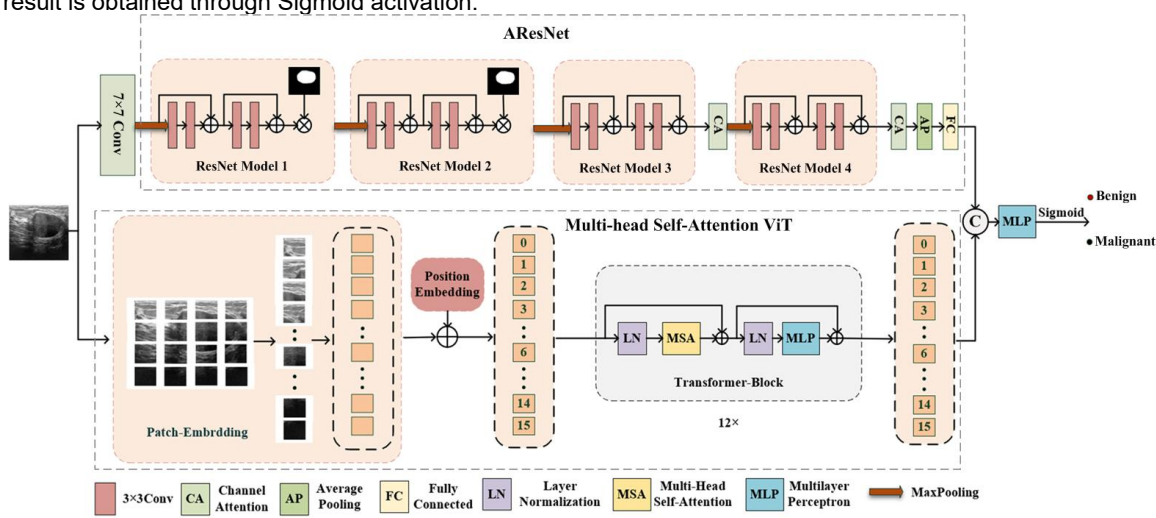

Figure 1: The network architecture of AResNet-ViT for ultrasound breast lesion classification

### 2.1 Local feature extraction

To enhance the network's capacity for focusing on and learning internal features of ultrasound breast nodules, we propose a locally-guided attention-based residual network named AResNet as the local feature extraction branch. This architecture is constructed based on the ResNet18 framework and comprises four residual blocks, each incorporating attention mechanisms, as illustrated in Figure 1. Within the structure of residual blocks 1 and 2, the network emphasizes intricate details, such as texture and edges, present in the ultrasound images. Given the substantial image size and the abundance of intricate details, the integration of spatial attention mechanisms becomes imperative to facilitate the network in effectively capturing and comprehending internal nodule information. The segmentation mask of the ultrasound breast nodule provides positional information and can act as a guide for spatial attention. Consequently, in residual blocks 1 and 2, we introduce ultrasound breast nodule segmentation mask attention (ROI-mask Attention, RA) [19].

The input features undergo two consecutive convolutions with a kernel size of 3x3 to extract local contextual information. Residual connections are introduced to expedite network optimization. Finally, the output features are element-wise multiplied with the segmentation mask image. Residual block 1 or 2 is formulated as follows.

$$Y(x) = (F(x) + x) \times R(x) \times C \quad (1)$$

Where $x$ represents the input, $F(x)$ represents the features learned through the convolution block, $R(x)$ represents the nodule mask feature map, $Y(x)$ represents the learning feature output under the attention guidance of the segmentation mask, and C is used to match the dimensions of the residual block and the segmentation mask map.

Residual blocks 3 and 4 further extract high-level semantic features based on the information derived from residual blocks 1 and 2. Each output channel within these blocks represents a distinct high-level semantic representation, contributing differently to the overall high-level semantics. Therefore, a Channel Attention (CA) module, as illustrated in Figure 2, is employed in residual blocks 3 and 4 to enhance the network's focus on channel outputs and amplify the informative channel representations. The CA module conducts global average pooling and global maximum pooling operations on the input feature map. The resulting one-dimensional feature vectors from both pooling operations are then combined and encoded using a multi-layer perceptron (MLP). Subsequently, the encoded result is subjected to a Sigmoid activation function to obtain a vector representing the weights assigned to each channel. This vector is then element-wise multiplied with the deep features of the input module. The primary objective of this module is to assign varying weights to each channel, thereby amplifying the channel-specific information that effectively captures the high-level semantic features exhibited by the ultrasound breast nodules.

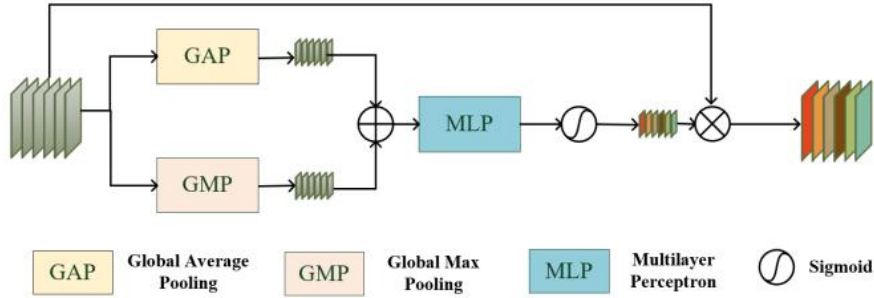

Figure 2: Channel Attention module

## 2.2 Global feature extraction

Convolutional neural networks (CNNs) primarily emphasize local receptive fields for information filtering while neglecting the global pixel-level self-correlation when processing ultrasound breast images. To augment the network's capability to acquire comprehensive global contextual information, this study incorporates a Vision Transformer (ViT) network that leverages the multi-head self-attention mechanism.The ViT network extracts both global image features and pixel-level self-correlation, as depicted in the lower branch of Figure 1. The network comprises 12 Transformer blocks connected in series. Each Transformer block independently performs self-attention and feed-forward neural network operations to iteratively extract features from the input sequence. This design enables multiple iterations of self-attention and feature extraction at various levels, thereby enhancing the model's expressiveness and performance.

The process of global feature extraction is as follows: First, the input image of size 224x224 is divided into blocks of size 16x16. Each image patch is linearly mapped and converted into a one-dimensional vector, preserving the spatial information between patches by adding position encoding. The data with position encoding is then fed into the Transformer blocks for layer-by-layer operations to perform feature encoding. By incorporating the self-attention mechanism, the network can capture the inter-dependencies among different locations in the image, facilitating a comprehensive understanding of the overall image context. This, in turn, improves the network's ability to extract the overall features and correlations of ultrasound breast images.

## 2.3 Loss function and evaluation metrics

### 2.3.1 Loss function

Given that breast ultrasound image classification is a binary classification task, the Binary Cross Entropy (BCE) loss function is employed, as denoted by formula 2.

$$L_{BCE} = -\frac{1}{N}\sum_{i=1}^{N}\left[y_i \log(p_i) + (1-y_i)\log(1-p_i)\right] \quad (2)$$

Where N represents the total number of pixels in the input image. $y_i \in \{0,1\}$ represents the true label of the i-th pixel, where 1 represents the pixel corresponding to the positive class and 0 represents the negative class. $p_i \in \{0,1\}$ indicates the probability that pixel i is predicted to be a positive class.

## 3 RESULTS

## 3.1 Experimental settings

### 3.1.1 Image Dataset

We assessed the performance of AResNet-ViT using the publicly available BUSI dataset, which was acquired from two ultrasound devices (LOGIQ E9 and LOGIQ E9Agile) at Bachia Hospital[20]. The images in the dataset have dimensions of 500x500, encompassing a total of 780 images. Among these, 487 images contain benign nodules, 210 images contain malignant nodules, and 133 images depict normal breast tissue. All images have been meticulously annotated by experienced medical professionals. Given the primary focus of this study on breast lesion recognition, experiments were conducted exclusively employing benign and malignant images. To ensure a rigorous evaluation, these images were meticulously partitioned into training and testing sets, maintaining a well-balanced 8:2 ratio. To address the issue of overfitting due to limited data samples, data augmentation techniques, including horizontal flipping and rotations at 90°, 180°, and 270°, are employed on the training set images. This effectively increases the sample size by a factor of five. Subsequently, all augmented and original images are resized from their original 500x500 dimensions to 224x224.

### 3.1.2 Experimental environment

The experiments in this work were all performed using Python 3.7 under a Windows 10 operating system. All deep learning models were developed using the Keras framework. The specific workstation parameters were as follows: Intel Xeon(R) CPU E5-2680 v4 @ 2.40GHz x 56 processor, two NVIDIA GeForce RTX 2080Ti GPU graphics cards, each GPU having 11GB of video memory.

We set the training parameters of the Adaptive Moment Estimation optimizer (Adam) with a learning rate of 0.0001 and a batch size of 4. To prevent overfitting, an early stopping technique was employed. Specifically, if the loss function on the validation dataset failed to decrease for 20 consecutive iterations, the training was halted.

### 3.2   Evaluation Metrics

The performance of all experiments was evaluated using accuracy rate (ACC), true positive rate (TPR), true negative rate (TNR), and area under the curve (AUC). ACC provides an overall assessment of the model's classification performance. TPR represents the probability of correctly classifying a malignant nodule as malignant. TNR represents the probability of accurately labeling a benign nodule as benign. AUC measures the area under the Receiver Operating Characteristic (ROC) curve, with TPR plotted on the vertical axis and False Positive Rate (FPR) on the horizontal axis. The AUC value ranges between 0 and 1, where a higher value indicates better classification performance. These evaluation metrics were defined as formulas (3)-(5).

$$ACC = \frac{TP+TN}{TP+FP+FN+TN} \quad (3)$$

$$TPR = \frac{TP}{TP+FN} \quad (4)$$

$$TNR = \frac{TN}{TN+FP} \quad (5)$$

Where TP represents the number of pixels with real labels as breast lesions and classified as breast lesions; TN represents the number of pixels with real labels as non-breast lesions and classified as non-breast lesions; FP represents the number of pixels with real labels as non-breast lesions but classified as non-breast lesions. The number of pixels of breast lesions; FN represents the number of pixels whose true label is breast lesions and is classified as non-breast lesions.

### 3.3   Ablation experiments

#### 3.3.1  Effectiveness of the attention mechanism

To validate the rationality and effectiveness of the attention guidance module, five ablation experiments were conducted, and the corresponding results are presented in Table 1. "Network 1" refers to the ResNet18 network without any attention added. "Network 2" incorporates segmentation mask attention after the completion of the first two residual blocks of the ResNet18 network, while "Network 3" incorporates segmentation mask attention after the completion of the last two residual blocks. "Network 4" integrates segmentation mask attention after the completion of all residual blocks in the ResNet18 network. Lastly, "Network 5" extends "Network 2" by further adding channel attention after the completion of the last two residual blocks. The same set of parameters was used across all experiments.

Table 1: Ablation analysis of Attention-Guided Mechanisms

| Algorithm structure | ACC | TPR | TNR |
| --- | --- | --- | --- |
| network1 | 0.796 | 0.770 | 0.804 |
| network2 | 0.835 | 0.824 | 0.852 |
| network3 | 0.827 | 0.839 | 0.816 |
| network4 | 0.843 | 0.831 | 0.859 |
| network5 | 0.857 | 0.842 | 0.866 |

From the table, it can be observed that regardless of where the segmentation mask attention is added, all evaluation metrics demonstrate improvement compared to the original ResNet18. This indicates the effectiveness of incorporating segmentation mask attention. However, the impact of adding attention at different positions on classification performance varies. Notably, the lowest accuracy is observed when the deep layers of "Network 3" have segmentation mask attention added. This decrease in accuracy can be attributed to the negative impact of incorporating segmentation mask attention on the representation of abstract information in the deeper layers of ResNet18. In contrast, "Network 4" achieves higher accuracy by adding segmentation mask attention to every residual block. However, the improvement is minor, and the computational complexity of "Network 4" is higher. Therefore, in "Network 5," we adopt "Network 2" and additionally incorporate channel attention after the completion of the last two residual blocks. All the evaluation metrics of "Network 5" are superior to those of other groups, indicating that the combination of segmentation mask and channel attention in Network 5 is effective.

*3.3.2 Effectiveness of the dual-branch architecture*

To assess the performance of each individual branch, as well as the combined architecture of the dual-branch architecture in ultrasonic breast nodule classification, ablation experiments were conducted across four experimental groups. The first group exclusively utilized the ResNet18 network for classification, while the second group employed the ViT network to classify benign and malignant breast nodules. In the third group, the ResNetA network was employed, incorporating a segmentation mask attention mechanism at the end of the first two residual blocks of the ResNet18 network, and channel attention at the end of the last two residual blocks, to conduct breast nodule classification experiments. The fourth and fifth groups were based on the ViT network architecture and involved the parallel fusion of the ResNet network and the AResNet network to classify benign and malignant breast nodules. The results of the ablation experiment are presented in Table 2.

As observed from the Table 2, the performance metrics of a single network (ResNet18 or ViT) were inferior to the combination of ResNet18 and ViT (ResNetViT network), signifying that network integration enables the learning of more relevant features specific to breast nodules. Furthermore, in comparison to the ResNet18 classification network, the AResNet classification network exhibited improvements in accuracy (ACC), true positive rate (TPR), true negative rate (TNR), and area under the curve (AUC), with respective increments of 0.061, 0.072, 0.062, and 0.066. This suggests that the AResNet classification network outperforms in guiding and learning nodule region features. Additionally, the parallel fusion of AResNet and ViT further enhanced performance metrics, with the most significant improvement observed in TNR. This indicates a higher recognition capability of AResNet-ViT model in identifying samples that manifest malignant nodules but are actually benign, which are crucial for clinical diagnosis as they represent the most easily misjudged cases.

Table 2: Ablation analysis of the dual-branch architecture

| Method | ACC | TPR | TNR | AUC |
|---|---|---|---|---|
| ResNet18 | 0.796 | 0.770 | 0.804 | 0.813 |
| ViT | 0.835 | 0.815 | 0.842 | 0.851 |
| AResNet | 0.857 | 0.842 | 0.866 | 0.879 |
| ResNet-ViT | 0.863 | 0.855 | 0.874 | 0.903 |
| AResNet-ViT | 0.889 | 0.861 | 0.896 | 0.925 |

### 3.4 The heat-maps of classification results

The visual classification results for the test samples obtained from the AResNet-ViT model are depicted in Figure 3, in which the original ultrasound breast nodule images are displayed in the top row, while the feature attention heat map generated by the AResNet-ViT model is presented in the bottom row. The feature attention heat map assigns weights to each position of the input data, indicating the areas or features to which the model pays greater attention. This visualization enables us to identify the specific regions within the input data that the model finds most significant. It can be observed from the figure that the nodule area receives primary attention from the model, as indicated by the highly weighted regions in the heat map.

Furthermore, in breast ultrasound images where the internal ultrasound characteristics of the nodule resemble those of the surrounding tissue, the model demonstrates accurate discrimination between the nodule area and the background. Additionally, the prediction results of the AResNet-ViT model for benign and malignant nodule samples with overlapping representations were found to align with the gold standard, indicating the model's capability to achieve precise classification.

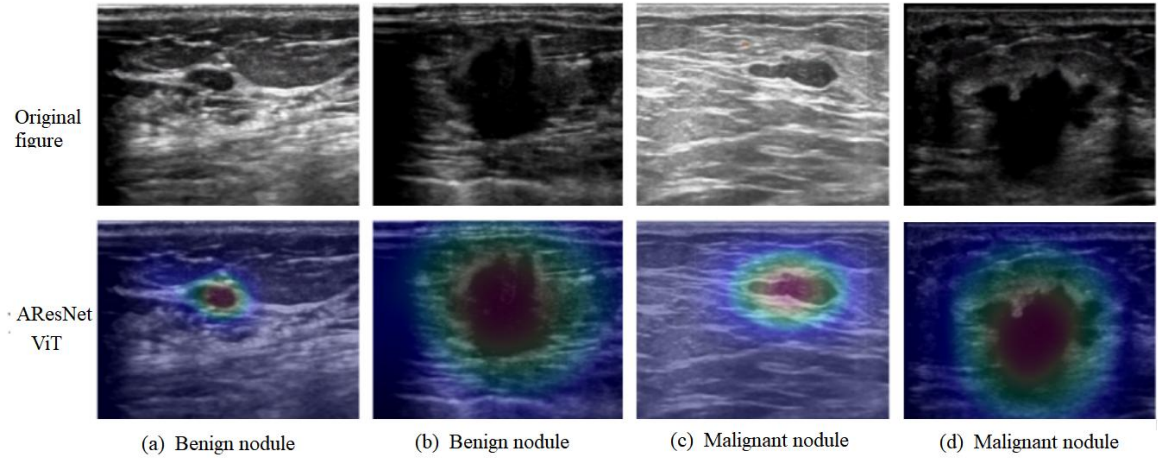

(a) Benign nodule  (b) Benign nodule  (c) Malignant nodule  (d) Malignant nodule

Figure 3: Heat map of the Classification Results

## 3.5 Comparative analysis

to investigate whether AResNet-ViT outperforms the existing classical models and other methods published in the field, we conducted a comparative analysis. The analysis is divided into two parts: an initial comparison with four well-established classical models (VGG16 [21], ResNet34 [22], DenseNet [23], and InceptionV3 [24]), followed by a comparison with three recently published methods. except for the dataset used in reference [25], all other methods, including the one proposed in the study, utilize the same BUSI dataset.

Table 3: Performance metrics of the compared methods

| Method | ACC | TPR | TNR | AUC | Data Sources |
|---|---|---|---|---|---|
| VGG16 | 0.782 | 0.749 | 0.789 | 0.791 | BUSI |
| ResNet18 | 0.796 | 0.770 | 0.804 | 0.813 | BUSI |
| DenseNet | 0.827 | 0.801 | 0.842 | 0.847 | BUSI |
| InceptionV3 | 0.824 | 0.817 | 0.831 | 0.852 | BUSI |
| literature[20] | 0.758 | 0.801 | - | - | private dataset |
| literature[21] | 0.855 | 0.707 | 0.827 | 0.865 | BUSI |
| literature[22] | 0.869 | 0.869 | - | 0.912 | BUSI |
| Our | 0.889 | 0.861 | 0.896 | 0.925 | BUSI |

From the first four rows of table 3, it is evident that our model achieves the most outstanding classification results compared to the classical models, indicating that our model outperforms the classical models in terms of accuracy in predicting the benign or malignant nature of nodules. Specifically, our classification model demonstrates a higher True Positive Rate (TPR), indicating its ability to identify more lesion areas and a lower rate of missed diagnoses. Moreover, our model exhibits a higher True Negative Predictive (TNP) value compared to the classical models, ranging from 0.054 to 0.107. This suggests that our model demonstrates higher accuracy in classifying challenging samples that exhibit malignant nodule characteristics but are actually benign.

The comparison with other literature is presented in rows 5-7 of Table 3. Reference [26] introduces additional embedding methods based on Transformer networks to enhance classification performance, but our method demonstrates superior performance across all metrics. Reference [27] employs a dual-channel input, extracting and merging features from ultrasound breast nodule images and breast X-ray images of different modalities. While it slightly outperforms our method in terms of True Positive Rate (TPR) and Area Under the Curve (AUC), its accuracy (ACC) in nodule classification is lower. Furthermore, the metrics in reference [25] also exhibit inferior performance compared to our method, with the paper itself acknowledging unsatisfactory results in classifying challenging samples with similar benign and malignant representations. In summary, our proposed AResNet-ViT network achieves the highest performance in terms of Accuracy (ACC), TPR, True Negative Rate (TNR), and AUC among the four evaluated metrics.

## 4 DISCUSSION

In this study, a hybrid CNN-Transformer architecture called AResNet-ViT was proposed for the benign-malignant classification of breast nodules in breast ultrasound images. The AResNet-ViT model combines the ability of CNNs to extract localized features and the capability of Transformers to model global features, resulting in a more discriminative feature representation for accurate classification. AResNet-ViT is designed with a dual-branch architecture. One branch focuses on extracting local detailed features from images using a residual network based on the ResNet18 framework. This branch consists of four residual blocks, each of which incorporates an attention mechanism. The other branch leverages a Vision Transformer (ViT) for global feature extraction. We use segmentation mask attention and channel attention for the shallow and deep modules of residual networks respectively, because the shallow layers of residual networks mainly extract low-level semantic features, paying more attention to nodule location information, whereas the deep residual networks extract high-level semantic features, and the weight of channels is more important than nodule location. Ablation experiments conducted on the residual network verify that the combined use of both types of attention achieves higher evaluation metrics compared to using segmentation mask attention or channel attention alone.

Ablation experiments were conducted to evaluate the performance of different architectures in ultrasonic breast nodule classification. The study compared ResNet18, ViT, AResNet (ResNet18 with segmentation mask and channel attention), and the fusion of AResNet and ViT. The results showed that the combination of ResNet18 and ViT (ResNetViT) outperformed individual networks, indicating the benefit of network integration. AResNet exhibited improvements over ResNet18 in accuracy, true positive rate, true negative rate, and area under the curve. The fusion of AResNet and ViT further enhanced performance, particularly in identifying samples with malignant characteristics but are actually benign, which are critical for accurate clinical diagnosis.

The heat maps of the classification results demonstrate that the AResNet-ViT model can accurately discriminate between the nodule area and the background, even in cases where the internal ultrasound characteristics of the nodule resemble those of the surrounding tissue. This further confirms the AResNet-ViT model's ability to learn and recognize features specific to the nodule region, indicating its precise classification capability. When compared to classical models and recently published methods in the field of ultrasonic breast nodule classification, our AResNet-ViT model demonstrates superior performance across all metrics, including ACC, TPR, TNR, and AUC of 0.889, 0.861, 0.896, and 0.925, respectively. The results indicate that the CNN-Transformer hybrid architectures can significantly improve ultrasound breast nodule classification. Moreover, integrating attention mechanisms in the convolutional stages enhances the extraction of local features.

Despite its outstanding performance, our method had some limitations, such as the requirement for a large dataset for effective training. Breast ultrasound images are intricate and vary greatly between individuals, making it challenging to build a robust classifier with limited data. Future work should focus on collecting a larger and more diverse dataset to improve model generalization. In addition, the computational complexity of the hybrid model can be high, making it challenging to deploy in real-time clinical settings. Therefore，future work should also optimize the model for efficient computation and reducing inference time is essential for practical applications.